\documentclass[twocolumn,aps, prl, showpacs]{revtex4}  
\usepackage{graphicx}

\begin{document}

\title{Scale-dependent rigidity of polymer-ornamented membranes}
\author{Thomas Bickel}
\altaffiliation{Present address: Physics Department, University of California, Los Angeles, CA 90024, USA. \\
Email address: bickel@physics.ucla.edu} 
\author{Carlos M. Marques}
\affiliation{Laboratoire de Dynamique des Fluides Complexes, CNRS-ULP, 67084 Strasbourg, France}

\begin{abstract}
  We study the fluctuation spectrum of fluid membranes carrying grafted
polymers. Contrary to usual descriptions, we find that the
modifications induced by the polymers cannot be reduced to the
renormalization of the membrane bending rigidity. Instead we show that
the ornamented membrane exhibits a scale-dependent elastic modulus that
we evaluate. In ornamented lamellar stacks, we further show that this
leads to a modification of the Caill\'e parameter characterizing the
power-law singularities of the Bragg peaks. \\

\end{abstract}


\maketitle

The characterization of polymer-membrane interactions is a fundamental issue of colloidal science. In cosmetics,
pharmaceutics or detergency, many formulations are suspensions of
self-assembled surfactant bilayers with
polymers added  for  performance,
processing, conditioning or delivery~\cite{vandepas94}. Likewise, lipid bilayers form the
walls of living cells and liposomes, and host a great variety of
macromolecules for coating protection, ion exchange and
mechanical reinforcement.~\cite{albertsbook}.  In many instances, the
polymers are end-tethered to the soft
interfaces. Grafting is easily achieved experimentally by using polymer
chains that carry hydrophobic groups. Typical examples are provided by the
so-called PEG-lipids, hydrophilic chains of
polyethylene glycol covalently linked to a double-tail phospholipid
molecule. Recent studies have
shown that polymers grafted to bilayers can induce
gelation~\cite{warriner96} or other phase
changes~\cite{yang98,castro99} in liquid lamellar phases. They stabilize
monodisperse vesicles~\cite{joannic97} and
modify the geometry of monolamellar~\cite{ringsdorf91,cates91} and
multilamellar~\cite{frette99} cylindrical
vesicles. They also lead to drastic changes in the structure and phase
behaviour of ternary amphiphilic
systems~\cite{endo00}.

As first explained by Canham and Helfrich, fluid membranes
are fluctuating objects~\cite{seifert97}. In thermal equilibrium, the surface assumes
all the possible
shapes allowed by the geometry with the associated Boltzmann
probability ${\cal P}\sim \exp\{-{\mathcal H}_0/(k_B T)\}$.
The Hamiltonian ${\mathcal H}_0$ is a quadratic function of $c_1$ and $c_2$,
the two principal curvatures at any given point of the surface, 
\begin{equation}
{\mathcal H}_0= \int dS\  \left\{ \frac{\kappa}{2}(c_1+c_2-2c_0)^2 +
\bar{\kappa }c_1c_2\right\}   \ .
\end{equation}
The spontaneous curvature $c_0$ vanishes for symetric
bilayers. The Gaussian rigidity
$\bar{\kappa }$ plays an important role in the determination of the
topology of the system, and the integral $ \int
dS\ \bar{\kappa }c_1c_2 $ does not depend on the particular shape of the membrane. The amplitude of the
thermal fluctuations is controled by the bending rigidity
$\kappa $, that ranges from a few $k_BT$ to a few tens of $k_BT$. Conversely, the analysis of the
height correlations in a membrane system allows for the determination of the
constitutive rigidity. Many techniques were specifically developped to extract $\kappa $ by
comparing experiments and theoretical
predictions for vesicles, lamellar stacks,
bicontinuous phases and other geometries.
These methods have so far been of limited application in polymer-membrane systems because of the lack of prediction for the fluctuation spectrum of bilayers in the presence of
polymers. In this Letter, we
make a first step to reach this gap by investigating the role of
grafted chains on the
fluctuation spectrum of fluid membranes. We first analyze the subtle interplay
between {\em local} monomer concentration and
surface curvature. We show that the chains induce
additional height correlations that can be interpreted as the
result of effective, scale-dependent elastic
coefficients. For the specific case of an ornamented lamellar phase
of membranes, we also determine the
Caill\'e parameter that characterizes the power-law singularities
around the Bragg peaks.

In a mean-field description, ornamented membranes are
assumed to have new parameters
$\kappa_{\mathit{eff }} = \kappa + \Delta \kappa $ and
$\bar{\kappa}_{\mathit{eff }} = \bar{\kappa } + \Delta
\bar{\kappa } $ different from those of the bare membrane. We consider 
${\mathcal N}$
polymers grafted to each side of a membrane of area ${\mathcal S}$, the chains being free to diffuse along the membrane.  The surface coverage $\sigma = {\mathcal
N}/{\mathcal S}$ is low enough to avoid polymer/polymer
interactions: $\sigma < \sigma^* \simeq
R_F^{-2}$, with $R_F$ the Flory radius of the chains. In this limit of
small density, the excess surface energy
 is proportional to $\sigma $ and one anticipates
corrections of the order of $\Delta
\kappa \sim \Delta \bar{\kappa }  \sim k_B T (\sigma R_F^2)$ in both good
and theta solvent. The effect of  solvent quality and polymer architecture resides in the numerical prefactors
of the variations. For Gaussian chains tethered by one extremity, the calculations can be carried out explicitly and one gets $\Delta \kappa
 =  \left( 1+ {\pi
\over 2} \right)k_BT
 \sigma R_g^2 $ and $\Delta \bar{\kappa }  =  -2k_BT \sigma R_g^2$~\cite{hiergeist96,marquesfournier},
with
$R_g = (Na^2/6)^{1/2}$ the radius of gyration of a Gaussian 
 chain of $N$ monomers of size $a$. At the overlap concentration $\sigma^* $, the
variations are of the order of the thermal energy $k_BT$. However, for most practical
situations, the chains can freely move along the membrane surface: the degrees of
freedom associated with the end positions are annealed, as opposed to the quenched
results of references~\cite{hiergeist96,marquesfournier}. A thorough study of the
system reveals that averaging the free energy over the anchor position is not
equivalent  to averaging the partition function itself. As a consequence, the response
to bending is lowered if one correctly accounts for the tendency of the chains to be
at the top of the membrane ondulation, leading to~\cite{bickelprep}   
\begin{eqnarray}
\label{deltak}
\Delta \kappa & = & k_BT \sigma R_g^2   \ ,\\
\Delta \bar{\kappa } & = & -2k_BT \sigma R_g^2  \ .
\end{eqnarray}
The reduction of the effective modulus is due to the linear coupling between 
positions and  spontaneous curvature and actually follows the general argument first
recognized by Leibler~\cite{leibler86}.

These mean-field results miss however a crucial feature, namely that the
polymer-membrane interactions strongly depend on the length scale.
  We first consider
the  partition function ${\mathcal Z}_N$ of a single Gaussian chain and construct a 
perturbative expansion
  in the limit of small surface deformations around the flat plane.
In absence of overhangs and inlets, the surface profile can be described by
a single-valued function $h({\bf r})$, where
${\bf r}$ spans the reference plane. We expand ${\mathcal Z}_N$ in a
series ${\mathcal Z}_N^{(0)} + {\mathcal
Z}_N^{(1)} + {\mathcal Z}_N^{(2)} + \ldots $, where ${\mathcal Z}_N^{(i)}$
is of order $h^i$ and ${\mathcal Z}_N^{(0)}(z) \approx
z/(\sqrt{\pi }R_g)$ denotes the partition function of a chain tethered at a small
distance $z$ of a flat, impenetrable wall. It
has been shown recently that the first-order term of the expansion is
related to the pressure field applied by the
grafted chain to the surface~\cite{breidenich00,bickel00}. At distance
$r=\sqrt{x^2+y^2}$ from the grafting point,
this entropic pressure decays with the scaling form
$p(r) \sim k_BTr^{-3}$ ($a \ll r \ll R_g$) in both good and theta solvents,
and then vanishes sharply beyond
$R_g$~\cite{bickel01}.

\begin{figure}
\includegraphics[width=8cm]{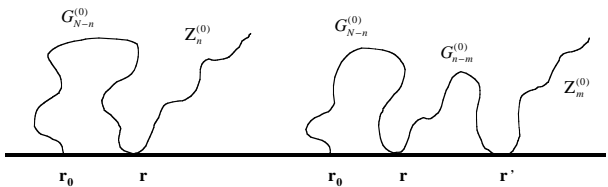}
\caption{Diagrammatic representation of the perturbative expansion of the
Gaussian chain partition function at first
order (left) and second order (right).}
\label{loop.eps}
\end{figure}

We now evaluate the contribution of the tethered polymers to the
fluctuation spectrum of the membrane. If one
 introduces the Fourier Transform (FT) of the height profile
$\tilde{h}({\bf q}) = \int d{\bf r}e^{-i{\bf qr}}
h({\bf r})$, then the energy ${\mathcal H}_0$ of the pure membrane reads
\begin{equation}
\label{fhelfrich}
{\mathcal H}_0 = {1 \over 2} \int
{ d{\bf q} \over (2\pi )^2 } \tilde{h}({\bf q}) \tilde{h}({\bf -q})\ \kappa
q^4  \ .
\end{equation}
Contrary to the previous mean-field results, we assert here that the 
presence of the chains entails non-local contributions to the elastic
description of the membrane.  This can be shown by extending the perturbative scheme up to second order
in $h$, a calculation that we perform
analytically for a Gaussian chain. In this case, the single-chain partition
function obeys the Edwards
equation~\cite{doibook}
\begin{equation}
\label{edwards}
\left( {\partial \over \partial n } - {a^2 \over 6 }\nabla^2 \right)
{\mathcal Z}_n ({\bf r},z) = 0  \ ,
\end{equation}
where $({\bf r},z)$ is the position of the constrained extremity. The
linearity of eq.~(\ref{edwards})
implies that each term of the development obeys an Edwards equation.
The solution of the successive orders are coupled through the
(Dirichlet) boundary condition on the wall ${\mathcal Z}_N({\bf 
r},h({\bf r})) =0$.
The solution of eq.~(\ref{edwards}) obeying the impenetrability condition
is then determined
recursively, and is given to each order by~\cite{bartonbook}
\begin{equation}
\label{partpol}
{\mathcal Z}_N^{(i)}({\bf r},z) = {a^2 \over 6} \int_0^N \!\!\! dn \int
d{\bf r}'
{\partial G_{N-n}^{(0)} \over
\partial z' }({\bf r},z ;{\bf r}',0)
{\mathcal Z}_n^{(i)}({\bf r}',0)  \ ,
\end{equation}
where the Green function $G_{N}^{(0)}$ of eq.~(\ref{edwards}) satisfies the
Dirichlet condition on the
horizontal plane~\cite{bickel01}. This solution corresponds to a ``loop
expansion'' of
${\mathcal Z}_N$: the term of order
$i$ enumerates the conformations that encounter the
surface $i$ times, as depicted on
figure~\ref{loop.eps}.  Finally, the statistical weight  ${\mathcal Z}_N$ is averaged for all possible positions of the anchoring
point along the surface.

\begin{figure} 
\includegraphics[width=8cm]{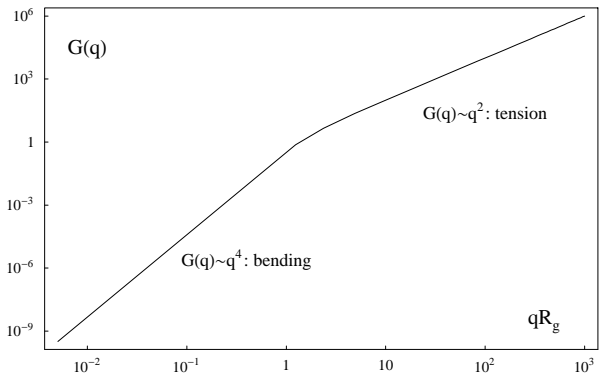}
\caption{Non-local polymer contribution to the fluctuation spectrum of a
membrane. At the natural length scale $R_g$,
the function $G(q)$ crosses over from a tension-like behaviour to a
bending-like behaviour.}
\label{correlations.eps}
\end{figure}

In general, the membrane carries a finite polymer concentration $\sigma $. We neglect
for sake of simplicity the interactions between the chains, whose origin is
twofold: in addition to the usual steric repulsions, the mean deformation
resulting from the polymer pressure gives rise to
attractive, deformation-sharing interactions~\cite{bickel01}.
However, the
latter do not modify the fluctuation spectrum of the
membrane. Our results are thus exact in the low-density limit $\sigma R_g^2 \ll
1$ where direct polymer interactions can be neglected.  Assuming $\mathcal{N}$ chains on each side of the membrane, the total partition function of the system is given by the path integral
\begin{eqnarray}
\mathcal{Z} & = & \frac{1}{(\mathcal{N}!)^2} \int \mathcal{D}[h] e^{-\beta \mathcal{H}_0} \left(\frac{1}{\mathcal{S}}\int d{\bf r} \mathcal{Z}_N\right)^{2\mathcal{N}}  \nonumber \\
& = &  \frac{1}{(\mathcal{N}!)^2}\int \mathcal{D}[h] e^{-\beta {\mathcal H}_{eff}}  \ ,
\end{eqnarray}
where the factor $1/\mathcal{N}!$ reflects the fact that particles are indistinguishable. The properties of the decorated
surface are described by the new Hamiltonian
$ {\mathcal H}_{eff} = {\mathcal H}_{0} + \Delta{\mathcal H}$, with
\begin{equation}
\label{helfrichdecore}
\Delta {\mathcal H}  =  {1 \over 2} \int
{ d{\bf q} \over (2\pi )^2 } \tilde{h}({\bf q}) \tilde{h}({\bf -q})G({\bf
q})  \ .
\end{equation}
The algebra will be detailed elsewhere~\cite{bickelprep}, and we now discuss the results. Integrating  out the degrees of freedom of the grafted chains generates the additional contribution
\begin{equation}
\label{correlation}
G({\bf q}) = {2k_BT  \sigma \over R_g^2 }\left\{ e^{-q^2R_g^2}
-1+q^2R_g^2 \right\}   \ .
\end{equation}
The correlation function $G({\bf q})$ is displayed on figure~\ref{correlations.eps}. At
length
scales larger than the polymer size $qR_g \ll 1$, one can  extract a 
correction to
the bending rigidity that reads $\Delta \kappa=\lim_{q \rightarrow 0} 
[q^{-4}G({\bf
q})] = k_BT  \sigma R_g^2$. We retrieve in this limit the mean-field correction~(\ref{deltak}), as expected. 
Interestingly, the polymer-membrane interactions also induce a
tension-like term at short distances: $G({\bf q})
\simeq 2 k_BT  \sigma q^2$ for $qR_g \gg 1$. The correlations between chain
and membrane conformations are
stronger inside the polymer mushroom: as a consequence, the fluctuations of
the surface are smoothed out at lenght
scales smaller than the polymer size. The coefficient of this effective
surface tension corresponds to the
two-dimensional perfect gas pressure (recall that {\em both sides} are
ornamented with ${\mathcal N}$ chains). Note
that our results contrast with the usual interfacial behaviour, where
tension is dominant at large scales whereas
bending effects appear only at shorter  scales. We should emphasize that
the results presented here do not depend on the nature of the
interface: they can be used to study a
wide class of surfaces, provided that
grafting of macromolecules is feasible.

We now apply our predictions to the lyotropic phase $L_{\alpha}$. This
phase consists in a stack of
regularly spaced membranes, that form a quasi-crystalline structure in one
direction while retaining their fluid
properties in the perpendicular plane. In pure water, the
smectic order is generally stabilized by long-range electrostatic
interactions. At
high salt concentration or for neutral surfactant molecules, the stability
of the phase arises from the entropic
repulsions between fluctuating membranes: we will refer to this system as
to the Helfrich system. In a continuum
elastic theory, the multilayer membrane system is described by the
displacement field
$u({\bf r})$ in the $z$ direction normal to the layers. The standard theory
of a pure $L_{\alpha}$ phase is based on
the smectic energy density~\cite{pggprostbook}
\begin{equation}
\label{smectic}
{F \over V}={1 \over 2} B \left( {\partial u \over \partial z }\right)^2
+{1 \over 2 } K \left(\nabla_{\perp }^2 u
\right)^2
\ ,
\end{equation}
the subscript $_{\perp}$ referring to the coordinates along the
layers. $K$ is the smectic curvature modulus and is directly related to the
membrane bending rigidity through the
relation $K=\kappa /d$, with $d$ the average layer spacing. The smectic
compressibility modulus $B$ reflects the
interlamellar interactions, and is given in a Helfrich system by $B =  9
\pi^2 (k_BT )^2 /(64 \kappa d^3)$.

The lyotropic phase is a good candidate to probe experimentally the fine
modifications to the properties of the pure phase induced by the addition
of macromolecules~\cite{tsapis01,sens01}.
X-ray or neutron scattering experiments allow to measure the exponent $\eta
$ that characterizes the power-law
singularities around the Bragg peak: $I(q) \sim \vert q-q_0
\vert^{-1+\eta}$, with $q_0 = 2\pi /d$~\cite{safinya86}. The ``strength'' of the smectic order varies conversely to $\eta $. For pure membrane sytems, the FT of the average heigth fluctuations is $\langle
\vert u({\bf q}) \vert^2 \rangle =  V k_BT
/( B q_z^2 + K q_{  \perp }^4 )$, which leads to the classical prediction
first derived by
Caill\'e~\cite{caille72}
\begin{equation}
\label{caille}
\eta = q_0^2 {k_BT \over 8 \pi \sqrt{KB}} \ .
\end{equation}

Likewise, a decorated membrane has according to eq.~(\ref{correlation}) a
scale-dependent bending rigidity
$\kappa ({\bf q}) = \kappa (1+ \alpha g_{\scriptscriptstyle D}(q^2R_g^2))$,
where we define the dimensionless
parameter $\alpha = {k_BT \over \kappa} \sigma R_g^2$, and
$g_{\scriptscriptstyle D}(x)={2 \over x^2}(e^{-x}-1+x)$ is
the Debye function~\cite{debye}.  This is certainly the most fundamental
step for forecasting the properties of
membrane-polymer complexes, but one should keep in mind that some
macromolecular interactions such as repulsion
between different membranes of the lamellar stack cannot be reduced only to
the modification of the single membrane
parameters. Notwithstanding this point, the height fluctuations are now
given by
\begin{eqnarray}
\label{hautdec}
\lefteqn{\langle \vert u({\bf q}) \vert^2 \rangle_{pm}  =  { V k_BT \over
\displaystyle B
q_z^2 + K q_{\perp }^4\left(1+\alpha g_{\scriptscriptstyle D}(q_{\perp
}^2R_g^2)\right) }} \nonumber \\
& & \simeq  \langle \vert u({\bf q}) \vert^2 \rangle - \alpha Kq_{\perp }^4
g_{\scriptscriptstyle D}(q_{\perp }^2R_g^2) {V k_BT  \over
\displaystyle \left( B q_z^2 + K q_{\perp }^4\right)^2  }   
\end{eqnarray}
at lowest order in $\alpha $. It ensues that the $q_{\perp }$ corrections lead to the same power-law
divergence of the scattering intensity with new exponent
\begin{equation}
\label{etadec}
\eta_{pm}=\eta \left( 1-{k_BT \over 2 \kappa } \sigma R_g^2  \right)  \ .
\end{equation}
The low-$q$ part of $G({\bf q})$ (\emph{i.e.}, the correction to $\kappa$) actually provides the main contribution to the integrals leading to~(\ref{etadec}). Providing that the interlamellar distance is not
affected, the tethered chains stiffen the lamellar order, that leads to a
sharpening of the Bragg peak. A decrease of $\eta$ with
polymer concentration is indeed observed in numerous experiments --- see for
instance~\cite{warriner96,castro99}. However, the measured variation of $\eta$
is much sharper than the predicted linear behaviour at low coverage. As
explained above, the chains are also likely to induce non-trivial
variations of the compressibility $B$. A simple explanation of these 
modifications has been
proposed by Castro-Roman \emph{et al.}~\cite{castro99}: a reduction in 
$\eta$ would be
due to the polymer layer that effectively increases the membrane thickness. Within this picture, however, it is still difficult to 
explain why
the membrane thickness varies so sharply with polymer coverage. Our 
study suggests that
 this could be caused by a non-homogeneous chain distribution, arising from
the coupling between position and curvature. This naturally leads to a higher
density of polymers in curved regions, which also correspond to the
contact zones between the fluctuating bilayers. The concentration at which
$\eta$ is significantly modified should then be as lower as the 
coverage needed to
roughly have one polymer by membrane contact point.

In conclusion, we have developed a local description of the 
fluctuation spectrum of
fluid membranes ornamented with grafted polymers. Contrary to current 
mean-field
predictions, we find that the effect of the grafted chains cannot be 
reduced to an
effective increase of the membrane bending rigidity. Instead, we show that the elastic coefficient is a scale-dependent quantity. On scales much larger than 
the polymer size,
the bending rigidity is indeed renormalized. This low-$q$ 
correction is
the dominant contribution to the so-called Caill\'e
exponent measured by X-ray experiments in L$_\alpha$ phases. On 
scales smaller than
the polymer size, polymer-membrane interactions smooth the fluctuations,
leading to a tension-like contribution to the correlation function. Our results
were obtained for Gaussian chains, a good representation for polymers in
$\theta$-solvents. It is not clear at this point whether or not this 
description is robust
with respect to the introduction of monomer-monomer  excluded volume 
interactions.  In
particular, the ``tension'' regime that develops at short length
scales might exhibit a different, non-trivial form.
Indeed, the Debye function
$g_{\scriptscriptstyle D}$ is known to scale as
$g_{\scriptscriptstyle D}
\sim q^{-1/\nu }$, with $\nu =1/2$ in theta solvent and $\nu \simeq 3/5$ in good
solvent conditions. If the relation $G
\propto q^4g_{\scriptscriptstyle D}$ still holds for chains with excluded
volume, one should then expect $G({\bf q})$
to behave like $ G({\bf q}) \sim q^{7/3}$ for large wavevectors ${\bf q}$.
Work on this issue is currently under progress.

\acknowledgments
We gratefully acknowledge E. Blokhuis, R. Bruinsma,
A. Lau and P. Sens for inspiring
discussions. This work was supported by the Chemistry Department of
the CNRS, under AIP ``Soutien aux Jeunes
Equipes''.


\begin{thebibliography}{99}
\bibitem{debye}
The Debye function corresponds, up to a factor $N$, to the stucture factor
of a Gaussian chain. Note
that in our case $g_{\scriptscriptstyle D}$ is evaluated for $q_z=0$, that
corresponds to an average over the
$z$-direction in direct space.
\end{thebibliography}
\end{document}